\documentclass[twocolumn,pre,showpacs,10pt]{revtex4-1}
\usepackage{graphicx}
\usepackage{amsmath}
\usepackage{amsfonts}
\usepackage{mathtools}
\usepackage{subfigure}
\usepackage{mhchem}

\begin{document}
\newcommand{\nwc}{\newcommand}
\nwc{\vs}{\vspace}
\nwc{\hs}{\hspace}
\nwc{\la}{\langle}
\nwc{\ra}{\rangle}
\nwc{\lw}{\linewidth}
\nwc{\nn}{\nonumber}
\nwc{\pd}[2]{\frac{\partial #1}{\partial #2}}
\newcommand{\be}{\begin{equation}}
\newcommand{\ee}{\end{equation}}
\newcommand{\bea}{\begin{eqnarray}}
\newcommand{\eea}{\end{eqnarray}}

\title{Precision and dissipation of a stochastic Turing pattern}

\author{Shubhashis Rana and Andre C Barato}
\affiliation{Department of Physics, University of Houston, Houston, Texas 77204, USA}

\parskip 1mm
\def\d{{\rm d}}
\def\Ps{{P_{\scriptscriptstyle \hspace{-0.3mm} s}}}
\def\MF{{\mbox{\tiny \rm \hspace{-0.3mm} MF}}}
\def\ts{\tau_{\textrm{sig}}}
\def\tos{\tau_{\textrm{osc}}}
\begin{abstract}
Spontaneous pattern formation is a fundamental scientific problem that has received 
much attention since the seminal theoretical work of Turing on reaction-diffusion
systems. In molecular biophysics, this phenomena often takes place under 
the influence of large fluctuations. It is then natural to 
inquire about the precision of such pattern. In particular, spontaneous 
pattern formation is a nonequilibrium phenomenon, and the relation between the 
precision of a pattern and the thermodynamic cost associated with it remains 
unexplored. Here, we analyze this relation with a paradigmatic stochastic 
reaction-diffusion model, the Brusselator in one spatial dimension.   
We find that the precision of the pattern is maximized for an intermediate 
thermodynamic cost, i.e., increasing the thermodynamic cost beyond this value
makes the pattern less precise. Even though fluctuations get less pronounced   
with an increase in thermodynamic cost, we argue that larger fluctuations can  
also have a positive effect on the precision of the pattern. 
\end{abstract}
% Explanation of PACS numbers:
% 05.70.Ln: Nonequilibrium and irreversible thermodynamics
% 02.50.Ey: Stochastic processes 

\maketitle

%====================================================================================================================================================
\section{Introduction}
%====================================================================================================================================================

The formation of patterns is a ubiquitous phenomenon in nature, with embryogenesis 
as a prominent example. In 1952, Alan Turing introduced a fundamental concept in pattern formation, 
he showed that a homogeneous reaction-diffusion system can form a periodic spatial profile \cite{turi52}. 
The counter-intuitive idea introduced by Turing was that diffusion can generate an instability that
leads to spontaneous pattern formation. Turing patterns have been the subject of several theoretical 
and experimental works \cite{murr03,kond10,bres14}.  

Reaction-diffusion models that display spontaneous pattern formation are  traditionally described by 
deterministic nonlinear partial differential equations. However, the formation of patterns often takes
place in a setup where fluctuations are relevant, which requires reaction-diffusion models that 
take noise into account for an accurate description. For instance, the range of parameters for which  
a pattern is formed can be substantially increased in a stochastic reaction-diffusion model as 
compared to its deterministic version \cite{lugo08,butl09,bian10,butl11,wool11,schu13,bian17}.
Interestingly, this concept that noise increases the region in parameter space that leads to 
the formation of a pattern has been recently verified in experiments \cite{patt18,kari18}.          

A different question about the formation of a pattern in a stochastic setting is: how 
precise is the pattern? In particular, pattern formation happens in nonequilibrium 
systems that dissipate energy. What is the relation between the precision of  
the pattern and the thermodynamic cost to maintain it? In this paper, we address these questions.

They are part of larger research program that has been carried out recently, 
which is the study of the relation between precision and dissipation in biophysics 
\cite{qian07,lan12,meht12,gove14a,bara14a,bara15a,mcgr17,chiu19}. 
A problem in this context particularly connected with our work is the relation between 
the precision of temporal biochemical oscillations and energy dissipation 
\cite{qian00,cao15,bara17a,nguy18,fei18,wier18,mars19,zhan20,junc20a,junc20b,frit20}. Such biochemical oscillators are 
modeled as a system of chemical reactions without spatial structure. Here, we analyze a system 
of chemical reactions with spatial structure.    

We analyze the relation between precision and dissipation of a Turing pattern for 
a particular model, the Brusselator in one spatial dimension \cite{bian10}. We consider 
a thermodynamically consistent version of this model which falls within the 
framework of stochastic thermodynamics \cite{seif12}. This framework allows us to evaluate 
the rate of entropy production that quantifies the thermodynamic cost.
It is worth mentioning that the relation between stochastic thermodynamics and 
deterministic reaction-diffusion models has been established in \cite{fala18}.     

We find that increasing the thermodynamic cost does not necessarily improve the precision of a Turing pattern.
There is an optimal parameter value beyond which increasing thermodynamic 
cost reduces the  precision of the pattern. The interesting fact we explain in this paper is that 
fluctuations, which get less pronounced with the increase of the thermodynamic cost in our model, 
can have a positive effect in the precision of a Turing pattern. From a technical side, our results
are obtained with numerical simulations since we consider a regime for which known analytical 
approximations \cite{mack14} are not valid. 

The paper is organized in the following way. In Sec. \ref{sec2} we introduce the model and 
explain its basic phenomenology. We evaluate the rate of  entropy production
that quantifies the thermodynamic cost in Sec. \ref{sec3}. In Sec. \ref{sec4} we define the observables that quantify 
the precision of a pattern and evaluate them. We conclude in Sec. \ref{sec5}.

%====================================================================================================================================================
\section{The Brusselator in one spatial dimension}
%====================================================================================================================================================
\label{sec2}
%====================================================================================================================================================
\subsection{Chemical reactions in a single cell}
%====================================================================================================================================================

Our version of the Brusselator model in one dimension is defined in the following way. 
There are two chemical reactants, $X$ and $Y$. The system 
has $L$ cells labelled by the index $i=1,2,\ldots,L$. In each cell,  $X$ and $Y$ react according to the following scheme, 
\bea
\ce{A &&<=>[a_1][a_2] X_i},\nn\\
\ce{S + X_i &&<=>[b_1][b_2] Y_i + P},\nn\\
\ce{2X_i + Y_i&& <=>[c_1][c_2] 3X_i},
\label{eqscheme1}
\eea
where the chemical species $A$, $S$ and $P$ have fixed concentrations that are sustained by chemostats.
The parameters $a_1$, $a_2$, $b_1$, $b_2$, $c_1$, and $c_2$ are transition rates. These rates already 
account for the dependence on the concentrations of the chemical species $A$, $S$, and $P$. For example, the rate $b_1$
is proportional to the concentration of $S$.   

The thermodynamic force that drives the system of chemical reactions in Eq. \eqref{eqscheme1} out of equilibrium 
is the chemical potential difference $\Delta \mu=\mu_S-\mu_P$, where $\mu_S$ ($\mu_P$) is the chemical potential of the substrate $S$ (product $P$).
In order to establish a relation between $\Delta\mu$ and the transition rates we consider the following cycle with two chemical reactions: first 
the one with rate $b_1$ and then the one with rate $c_1$. After this cycle, the numbers molecules $X$ and $Y$ remain the same, however, a molecule $S$
is consumed and a molecule $P$ is produced. The generalized detailed balance relation \cite{seif12} is a postulate of stochastic thermodynamics that connects the 
thermodynamic force with transition rates. For this cycle, it reads 
\bea
\ln\frac{b_1c_1}{b_2c_2}=\frac{\Delta \mu}{k_BT},
\label{eqmu}
\eea
where $k_B$ is Boltzmann's constant and $T$ is the temperature of the external reservoir. 
We set $k_B=T=1$ throughout. If $\Delta \mu=0$ the system fulfills detailed balance and is at equilibrium. The chemical species $A$ 
cannot be consumed or produced in sequence of transition that leave the numbers of $X$ and $Y$ unaltered, hence, it is not related to 
a thermodynamic force that drives the system out of equilibrium.  

%====================================================================================================================================================
\subsection{Spatial diffusion and full model}
%====================================================================================================================================================

Besides the chemical reactions in Eq. \eqref{eqscheme1} the chemicals $X$ and $Y$ also diffuse 
between nearest neighbor cells. The transition from site $i$ to site $j=i\pm 1$ is represented 
by the scheme 
\bea
\ce{X_i ->[\alpha] X_j},\nn\\
\ce{Y_i ->[\beta] Y_j}.
\label{eqscheme2}
\eea
where $\alpha$ and $\beta$ are the diffusion rates of species $X$ and $Y$, respectively. We consider 
periodic boundary conditions throughout, in our notation, for $i=1$ the left nearest neighbor is $j=i-1=L$ and for $i=L$
the right nearest neighbor is $j=i+1=1$. We point out that boundary conditions can have an important role in the formation of 
Turing patterns \cite{klik18}.     

The number of $X$ ($Y$) molecules in site $i$ is denoted by $n_i$ ($m_i$). The state of the system is fully specified 
by the vectors $\vec{n}=\{n_1,n_2,\ldots,n_L\}$ and $\vec{m}=\{m_1,m_2,\ldots,m_L\}$. 
The transition rates associated with the reactions in Eq. \eqref{eqscheme1} are given by 
\bea
T_1(n_i + 1,m_i|n_i,m_i) &&=a_1,\nn\\
T_2(n_i - 1,m_i|n_i,m_i) &&=a_2\frac{n_i}{V},\nn\\
T_3(n_i-1,m_i+1|n_i,m_i) &&=b_1\frac{n_i}{V},\nn\\
T_4(n_i+1,m_i-1|n_i,m_i) &&=b_2\frac{m_i}{V},\nn\\
T_5(n_i + 1,m_i-1|n_i,m_i) &&=c_1\frac{n_i^2 m_i}{V^3},\nn\\
T_6(n_i-1,m_i+1|n_i,m_i) &&=c_2\frac{n_i}{V},
\label{eqrates1}
\eea
where $V$ represents the volume of the cell. In this notation $T_\nu(n_i',m_i'|n_i,m_i)$ is the transition 
rate from a configuration with $n_i$ and $m_i$ at site $i$ to a configuration with $n_i',m_i'$ at site $i$. The numbers 
$n_j$ and $m_j$ for all other $j\neq i$ are the same for both configurations. The transition rates associated with 
the diffusion transitions represented in Eq. \eqref{eqscheme2} are
\bea
T_7(n_i -1,n_j +1|n_i,n_j) = (\delta_{i,j-1}+\delta_{i,j+1})\alpha \frac{n_i}{2V},\nn\\
T_8(m_i -1,m_j +1|m_i,m_j) = (\delta_{i,j-1}+\delta_{i,j+1})\beta \frac{m_i}{2V}.
\label{eqrates2}
\eea
The time evolution of the probability of a configuration $(\vec{n},\vec{m})$ follows the 
master equation \cite{vankampen,mack14} with the transition rates given by Eq. \eqref{eqrates1} and Eq. \eqref{eqrates2}.

The parameters of the model are set to $a_1=1.5$,  $a_2=1.0$, $b_2=0.005$, $c_1=1.0$, $c_2=0.005$, $\alpha=1.0$, and $\beta=20.0$.
By varying the parameter $b_1$ we vary the thermodynamic force  $\Delta \mu$ in Eq. \eqref{eqmu}. We have performed numerical simulations 
using the Gillespie algorithm \cite{gill77}. The initial condition in our simulations is $n_i=700$ and $m_i=350$ for $i=1,2,\ldots,L$.
The volume $V$ is set to $V=500$. The system size is fixed as $L=200$. We point out that a standard analytical method to analyze 
stochastic reaction-diffusion models is 
the linear noise approximation \cite{mack14}. We resort to numerical simulations because this approximation does not work for parameter 
values that we are interested in, as explained below.  For the averages plotted here, we have performed $20$ independent realizations. In each 
realization we took $50$ different data points. The time interval between different data points in the same realization is $20$. 
 
\begin{figure}
\subfigure[]{\includegraphics[width=8cm]{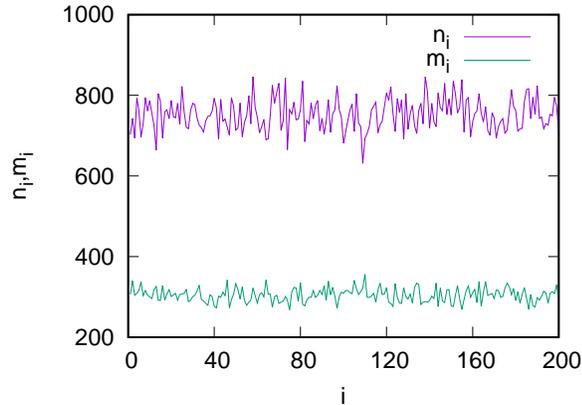}}
\subfigure[]{\includegraphics[width=8cm]{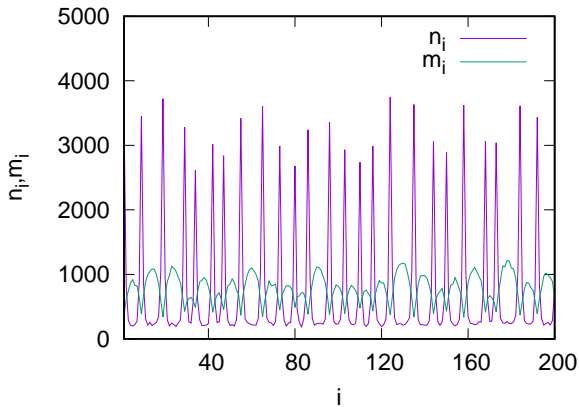}}
\vspace{-2mm}
\caption{(Color online) Spatial patterns. (a) Homogeneous profile  $\Delta\mu=10.5<\Delta\mu_c$. 
(b) Formation of pattern for $\Delta\mu=11.8>\Delta\mu_c$.
}   
\label{fig1} 
\end{figure}

%==========================================================================
\subsection{Onset of spatial oscillations}
%==========================================================================

\begin{figure}
\subfigure[]{\includegraphics[width=8cm]{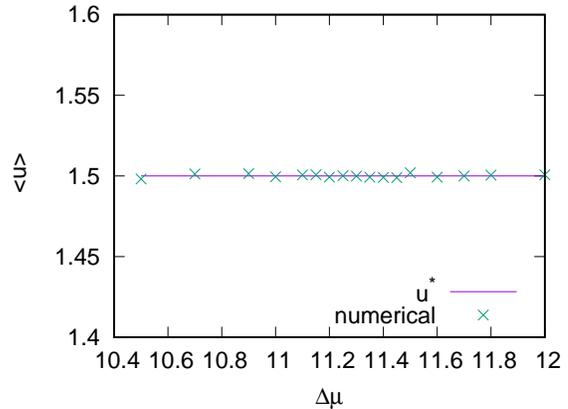}}
\subfigure[]{\includegraphics[width=8cm]{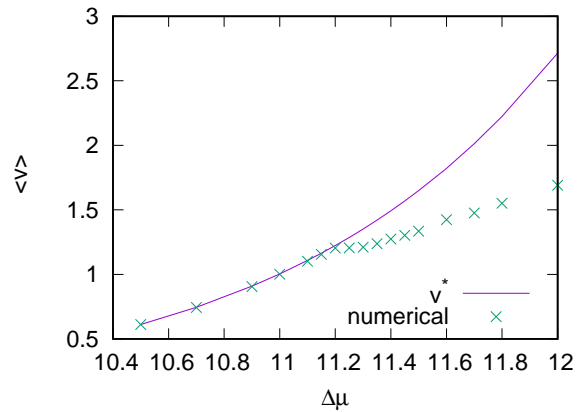}}
\vspace{-2mm}
\caption{(Color online) Average densities as functions of the thermodynamic force $\Delta\mu$.
The values for the homogeneous fixed point solution in Eq. \eqref{eqhomo}  are represented by the full magenta line.
}   
\label{fig2} 
\end{figure}

In the limit $L\to \infty$ this model is described by deterministic equations for the average concentrations $u_i\equiv n_i/V$ and $v_i\equiv m_i/V$. These equations can be obtained 
with standard methods \cite{mack14}, and they read
\bea
\frac{d u_i}{d\tau} &&=a_1 -a_2 u_i -b_1 u_i +b_2 v_i +c_1 u_i ^2 v_i -c_2 u_i^3 +\alpha \Delta u_i\nn\\
\frac{d v_i}{d\tau} &&=b_1 u_i -b_2 v_i -c_1 u_i ^2 v_i +c_2 u_i^3 +\beta \Delta v_i
\label{macro}
\eea
where $\Delta$ is a discrete Laplacian operator defined as $\Delta f_i= f_{i+1}+f_{i-1}-2f_{i}$.
The homogeneous fixed point of the above equation is
\begin{equation}
u^* = \frac{a_1}{a_2}\qquad v^* =\frac{b_1 u^* + c_2 (u^*)^3}{b_2 + c_1 (u^*)^2}.
\label{eqhomo}
\end{equation}
The emergence of a Turing pattern is related to the stability of this homogeneous solution. Depending on the values of the parameters
this solution may become unstable and an inhomogeneous one becomes stable. One of the conditions on the parameters for the formation of 
a pattern is that $\beta>\alpha$ \cite{murr03}. 

As shown in Fig. \ref{fig1}, for large enough $\Delta \mu$, an oscillatory pattern is formed. Both chemical species oscillate, with the minimums of 
the concentration $X$ at the same positions as the maximums of the concentration of $Y$, and vice versa. We can estimate the critical point $\Delta \mu_c\approx 11.1$ by 
analyzing the average densities 
\begin{equation}
\langle u\rangle\equiv \langle L^{-1}\sum_{i=1}^L u_i \rangle\qquad \langle v\rangle \equiv \langle L^{-1} \sum_{i=1}^L v_i\rangle,
\label{equv}
\end{equation}
where the brackets indicate an ensemble average.  This result is show in Fig. \ref{fig2}. Above the critical point the average concentration of $Y$ differs from the 
one obtained with the homogeneous fixed point.

\begin{figure}
\subfigure[]{\includegraphics[width=8cm]{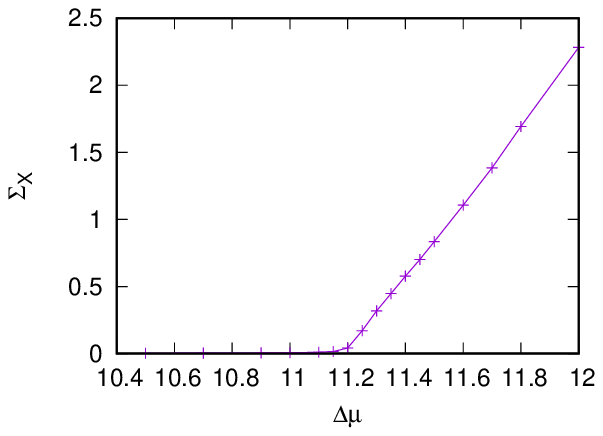}}
\subfigure[]{\includegraphics[width=8cm]{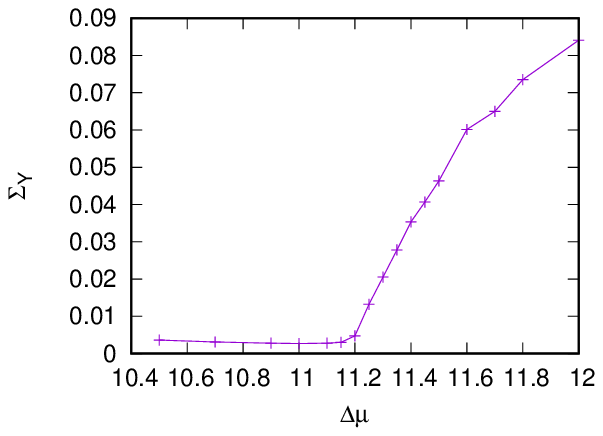}}
\vspace{-2mm}
\caption{(Color online) Relative standard deviations of the densities $\Sigma_X$ and $\Sigma_Y$  as functions of $\Delta \mu$.
The crosses represent the values obtained from numerical simulations and the line is a guide to the eye.
}   
\label{fig3} 
\end{figure}

Fluctuations of the densities are quantified by
\begin{equation}
\Sigma_X=\frac{\la (u-\la u \ra)^2\ra }{ \la u \ra^2}\qquad \Sigma_Y=\frac{\la (v-\la v \ra)^2\ra}{ \la v \ra^2}.
\label{eqfluc}
\end{equation}
As shown in Fig. \ref{fig3}, these quantities increase sharply above the critical point. These fluctuations do not quantify
the precision of the pattern but rather their increase with $\Delta\mu$ is simply related to the fact that the square of the deviation from the 
mean is typically much larger for a periodic profile than for a flat profile. In the next two sections we analyze the relation between precision 
and dissipation of a pattern in the regime $\Delta\mu>\Delta\mu_c$. Due to the large size of $\Sigma_X$ in this regime, 
we cannot use the linear noise approximation.

%==========================================================================
\section{Entropy production}
%==========================================================================
\label{sec3}

\begin{figure}
\includegraphics[width=8cm]{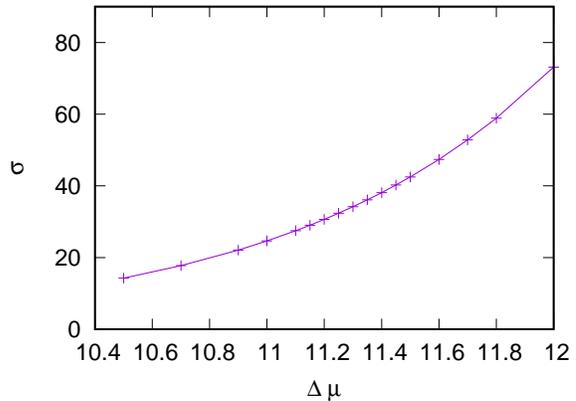}
\vspace{-2mm}
\caption{(Color online) Average rate of entropy production  $\sigma$ as a function of $\Delta\mu$.
The crosses represent the values obtained from numerical simulations and the line is a guide to the eye.
}   
\label{fig4} 
\end{figure}

For a stochastic trajectory of duration $T$, the stochastic rate of consumption of the substrate $S$ at cell $i$ is given by $J_i=\theta_i/T$,
where $\theta_i$ is a random variable that increases (decreases) by one whenever the chemical reaction 
$S + X_i\to  Y_i+P$ ($Y_i+P\to S + X_i$) takes place. It is straightforward to evaluate $J_i$ in a numerical simulation, by simply changing 
$\theta_i$ accordingly whenever these reactions take place.  The average rate of entropy production per cell is 
\begin{equation}
\sigma= \Delta\mu L^{-1}\sum_{i=1}^{L}\langle J_i\rangle.
\label{eqsigma}
\end{equation}
In words, the rate of entropy production is the rate of substrate consumption multiplied by the chemical potential difference of transforming
a substrate molecule $S$ into a product molecule $P$. Diffusion reactions do not show up explicitly  in this 
expression for $\sigma$ since the ratio of a diffusion transition and 
its reverse is one \cite{seif12}. However, $\sigma$ does depend on 
diffusion rates since $\langle J_i\rangle$ depends on these rates. 

In Fig. \ref{fig4} we plot $\sigma$ as a function of $\Delta\mu$. We observe that the rate of entropy production is an increasing function 
of $\Delta\mu$, the larger the thermodynamic force the larger the thermodynamic cost. In the next section we analyze the precision of the 
pattern as a function of $\Delta\mu$. Since the thermodynamic cost quantified by $\sigma$ is an increasing function of $\Delta\mu$, we refer 
to increasing (decreasing) $\Delta\mu$ as increasing (decreasing) the thermodynamic cost.  

\begin{figure}
\includegraphics[width=8cm]{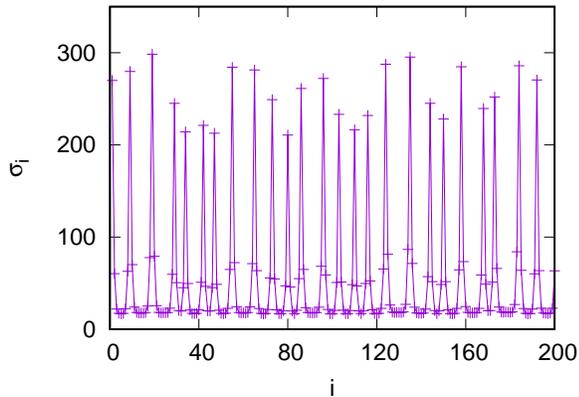}
\vspace{-2mm}
\caption{(Color online) Spatial profile of the stochastic rate of entropy production  $\sigma_i$ for $\Delta\mu=11.8$.
}   
\label{fig5} 
\end{figure}

Interestingly, there is a spatial profile of the entropy production if we consider a single stochastic trajectory, as shown in Fig. \ref{fig5}. The quantity $\sigma_i=\Delta\mu J_i$, which is the 
rate of entropy production at site $i$, displays spatial oscillations. The peaks of $\sigma_i$ coincide with the peaks of $u_i$, since 
a larger density of $X$ molecules increases the likelihood of the chemical reaction $S + X_i\to  Y_i+P$.

The critical behavior of the average rate of entropy production has been investigated in several models \cite{croc05,oliv11,tome12,bara12,zhan16,nguy18}.
We cannot identify any non-analytical behavior of rate of entropy production $\sigma$ or its first derivative with respect to $\Delta\mu$ within our numerics,
which does not discard non-analytical behavior of higher order derivatives.

%Nevertheless, we do observe a signature of the phase transition if we consider fluctuations of the entropy
%production, as quantified by the  the Fano factor 
%\begin{equation}
%F_\sigma= \frac{\langle S^2\rangle-\langle S\rangle^2}{\langle S\rangle} 
%\label{eqsigma}
%\end{equation}
%where $S\equiv\Delta\mu L^{-1}\sum_{i=1}^{L} J_i$. As shown in Fig. \ref{fig5}, $F_\sigma$ 

%In particular, the finite size scaling analysis 
%in Fig. \ref{fig5} shows that in the limit $L\to\infty$ the Fano factor $F_\sigma$ diverges at the critical point.
%This divergence of the Fano factor at criticality  has also been observed in other 
%nonequilibrium phase transitions \cite{nguy18}. Furthermore, the critical behavior 
%of the average rate of entropy production has been investigated in several models \cite{croc05,oliv11,tome12,bara12,zhan16}.

%==========================================================================
\section{Optimal Precision}
%==========================================================================
\label{sec4}

\subsection{The positive effect of fluctuations}

The pattern formed in a stochastic reaction-diffusion model fluctuates, 
in contrast to a pattern that is formed in the deterministic case.
We now analyze the precision of a stochastic pattern.

\begin{figure}
\includegraphics[width=8cm]{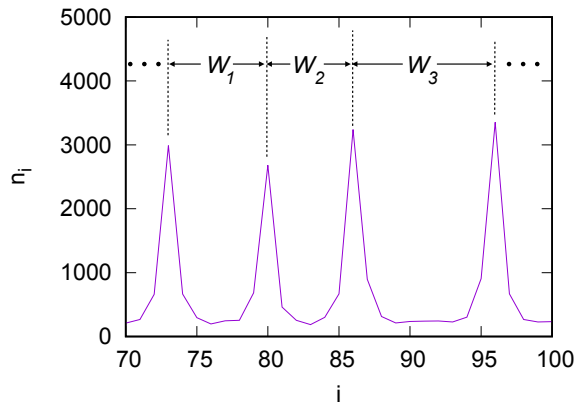}
\vspace{-2mm}
\caption{(Color online) Illustration of the variable $w$, which is the distance between two peaks.
For a given profile the variable $w$ is the sum of the distance between neighboring peaks 
divided by the total number of peaks. These distances are illustrated as $w_1$, $w_2$ and $w_3$ in the figure.
}   
\label{fig6} 
\end{figure}

In order to quantify the precision of the pattern  we consider the distance between two peaks $w$, which 
is illustrated in Fig. \ref{fig6}. We analyze the profile of the $X$ molecules since their peaks are more 
pronounced. Fluctuations of $w$ are quantified by
\begin{equation}
\Sigma_w= \frac{\langle w^2\rangle-\langle w\rangle^2}{\langle w\rangle^2}.
\end{equation}
The fluctuations in the distance between two peaks has been used to quantify the 
precision of temporal biochemical oscillations \cite{cao15}. 

\begin{figure}
\includegraphics[width=8cm]{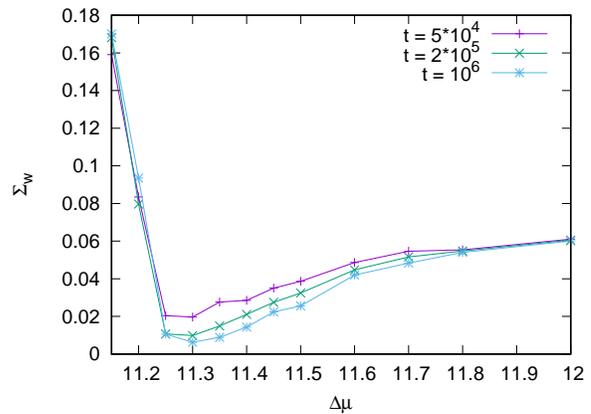}
\vspace{-2mm}
\caption{(Color online) Relation between precision of a pattern and thermodynamic cost. The relative standard deviation
$\Sigma_w$ as a function $\Delta\mu$. The lines are guides to the eye.}   
\label{fig7} 
\end{figure}

In Fig. \ref{fig7} we plot $\Sigma_w$ as a function of $\Delta\mu$. For the time 
$t=10^6$, this function seems to be close to saturation. Surprisingly,
this functions is not monotonically decreasing. There is an optimal value 
$\Delta\mu\approx 11.3$ for which $\Sigma_w$ is minimum. Hence, increasing 
the thermodynamic cost beyond this optimal value leads to a profile that is less precise.

A possible explanation for this result is as follows. If we watch a movie of the 
time evolution of a profile, we see that for larger $\Delta\mu$ the profile 
displays less pronounced fluctuations in the peak position, which remain static 
for longer periods of time for larger $\Delta\mu$. Hence, these fluctuations get 
less pronounced with an increase in thermodynamic cost. However, they can 
also be beneficial for the precision quantified by $\Sigma_w$. The pattern that is formed 
after some transient from the flat initial conditions depends on the particular stochastic trajectory. 
This pattern is different from the ``correct'' deterministic pattern. 
For large $\Delta\mu$ the system gets stuck in this pattern formed after the initial transient,
and fluctuations cannot alter the position of the peaks.

\begin{figure*}[t]
\subfigure[]{\includegraphics[width=59mm]{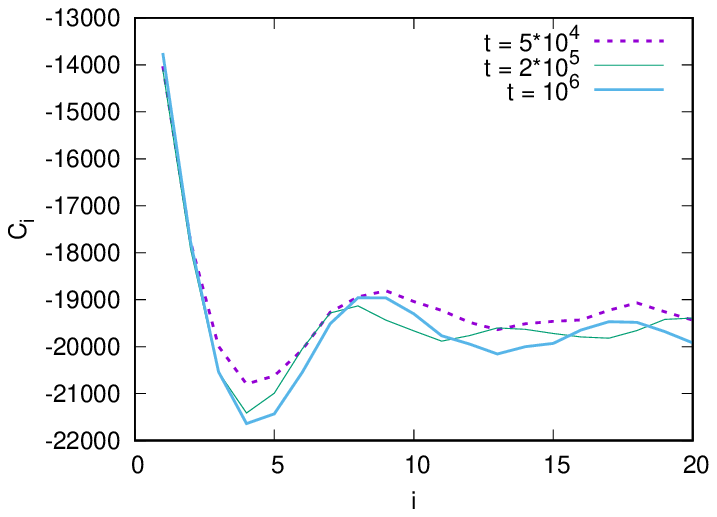}\label{fig8a}}
\subfigure[]{\includegraphics[width=59mm]{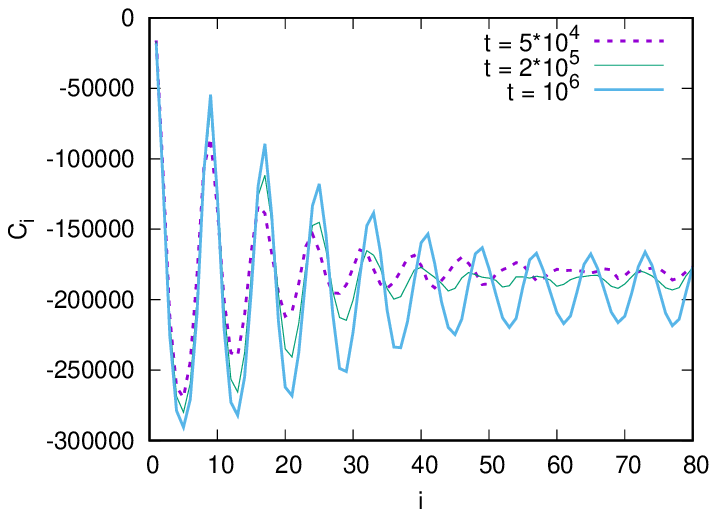}\label{fig8b}}
\subfigure[]{\includegraphics[width=59mm]{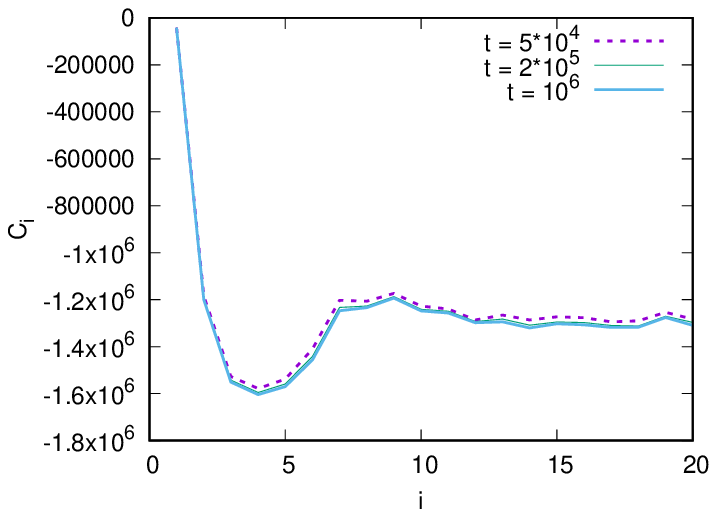}\label{fig8c}}
\vspace{-2mm}
\caption{(Color online) Spatial correlation $C_i$ for (a) $\Delta\mu=11$, (b) $\Delta\mu=11.3$ and, (c) $\Delta\mu=12$.
}   
\label{fig8} 
\end{figure*}

Fluctuations of the peak positions have then two competing effects on the precision of the pattern. 
The straightforward effect is that too much fluctuations destroy the precision of the 
pattern. The surprising effect is that fluctuations that are too small do not allow the
system to ``correct'' a sort of metastable pattern that is formed after the initial transient. 
The competition between these two effects leads to the optimal precision in Fig. \ref{fig7}.

The relation between precision and dissipation depends on the particular choice of 
how the transition rates depend on parameters such as $\Delta\mu$. While several models 
for temporal biochemical oscillations  do display a behavior for which  precision
improves with an increase in thermodynamic cost \cite{cao15}, examples for which 
precision of some kind is optimal at some intermediate thermodynamic cost do exist 
\cite{bara17a,nguy18,baie18,frit20}. The result for this particular model that could
be general, independent of the way transition rates depend on parameters, 
is that fluctuations also have a positive effect on the precision of a pattern
in a finite system within finite time.

\subsection{Spatial correlation function}

The random variable $w$ is hard to formally define with an equation (see \cite{mars19} for a discussion about this random variable concerning 
temporal oscillations). If we were to consider lower values of $\Delta\mu$, which are closer to the value of $\Delta\mu$ for which 
a pattern is first formed, it would be hard to differentiate between a peak and a fluctuation. 
A well defined observable that quantifies the precision of the pattern is the spatial correlation function 
\begin{equation}
C_i=\sum_{j=1}^L \left(\langle n_jn_{j+i}\rangle-\langle n_jn_j\rangle\right). 
\end{equation}

As shown in Fig. \ref{fig8}, this quantity displays oscillations that decay exponentially in space.
The number of coherent spatial oscillations, which is the correlation length divided by 
the average distance between two peaks, quantifies the precision of a pattern. This number of coherent 
oscillations is also used to quantify the precision of temporal oscillations \cite{cao15,bara17a}.
It is hard to get an exact value for this quantity and plot it as a function of $\Delta\mu$ within 
our numerics.  However, we do find clear evidence of our main result that precision is optimal 
for an intermediate thermodynamic cost. This evidence is shown in Fig. \ref{fig8}, where the spatial oscillations
in the correlation function are more robust for an intermediate  value of $\Delta\mu$.

%==========================================================================
\section{Conclusion}
%==========================================================================
\label{sec5}

We have analyzed the relation between precision of a spatial pattern and 
dissipation in a model that displays spontaneous pattern formation, the 
one dimensional Brusselator. A main result is that fluctuations  of the peak 
positions can also have a positive effect on the precision of the pattern, 
as they correct the ``metastable'' pattern that is formed after an initial 
transient of the stochastic dynamics. Such fluctuations also have the standard 
effect of decreasing the precision of the pattern. These competing effects 
leads to the result that the precision of the spatial pattern is optimal for 
an intermediate thermodynamic cost. 

Concerning future work we have considered a model with spatial oscillations 
and no temporal oscillations. It would be interesting to analyze the relation 
between temporal and spatial oscillations within models that display both 
oscillations, which are known as chemical waves \cite{anvi19}. 
Furthermore, the investigation of the effect of dimensions, reaction 
scheme, system size, and boundary conditions  on the relation between 
precision and thermodynamic cost of a Turing pattern will elucidate 
the degree of generality of the results found within the present 
model.

From a broader perspective, two universal results within stochastic thermodynamics 
concerning the relation between precision and thermodynamics cost are the 
thermodynamic uncertainty relation \cite{bara15a} and the bound on the number of coherent 
temporal oscillations obtained in \cite{bara17a}. The question whether a general relation
between spatial precision of some kind and thermodynamic cost exists remains an 
open one. Finally, it is an intriguing perspective to ask whether  biological systems 
operate close to the optimal value of the thermodynamic cost.

\bibliographystyle{apsrev4-1}

\bibliography{refs} 

\end{document}